\documentclass[a4paper]{article}

\usepackage{INTERSPEECH2022}
\usepackage{bm}
\usepackage{graphicx}
\usepackage{float}
\usepackage{ascmac} 
\usepackage{fancybox}
\usepackage{float}
\usepackage{tabularx}
\usepackage{booktabs}
\usepackage{amssymb}
\usepackage{multirow}
\usepackage{cite}
\usepackage{multirow}
\usepackage[table,xcdraw]{xcolor}
\usepackage[normalem]{ulem}
\useunder{\uline}{\ul}{}
\usepackage{url}

\def\L{{\cal L}}
\def\Lsdr{\L_{\text{sdr}}}

\def\Ln{\L_{n}}
\def\Lnd{\L_{n'}}
\def\Lcon{\L}
\def\WSDR{l}
\def\Lwh{\L_{\text{worst,hard}}}
\def\Lws{\L_{\text{worst,soft}}}
\def\Lsi{\L_{\text{multitask}}}
\def\Y{{\mathbf Y}}
\def\S{{\mathbf S}}
\def\W{{\mathbf W}}
\def\Shat{\hat{{\mathbf S}}}

\def\I{{\mathbf I}}

\def\N{{\mathbf N}}

\def\Cs{{\mathbf C}^{\text S}}
\def\Csc{\tilde{{\mathbf C}}^{\text S}}
\def\zs{{\mathbf e}^{\text S}}

\def\inRT{\in \mathbb{R}^{T}}
\def\SE{{\tt TSE}}

\def\SDR{{\tt SDR}}
\def\CE{{\tt CE}}
\def\softmax{{\tt Softmax}}
\def\one{{\mathbf l}^{\text S}}
\def\Extract{{\tt Extract}}
\def\Embed{{\tt Embed}}
\def\cands{\mathcal{C}}
\def\subcands{\mathcal{C'}}

\def\param{{\bm \Lambda}}
\def\parame{{\bm \theta_{\text{emb}}}}
\def\params{{\bm \theta_{\text{ext}}}}

\AtBeginDocument{
  \abovedisplayskip     =.6\abovedisplayskip
  \abovedisplayshortskip=.6\abovedisplayshortskip
  \belowdisplayskip     =.6\belowdisplayskip
  \belowdisplayshortskip=.6\belowdisplayshortskip}

\title{Strategies to Improve Robustness of Target Speech Extraction to\\ Enrollment Variations}

\name{Hiroshi Sato, Tsubasa Ochiai, Marc Delcroix, Keisuke Kinoshita, Takafumi Moriya, \\Naoki Makishima, Mana Ihori, Tomohiro Tanaka, Ryo Masumura}

\address{NTT Corporation, Japan}
\email{hiroshi.satou.bh@hco.ntt.co.jp}

\ninept

\begin{document}

\maketitle
\begin{abstract} 
Target speech extraction is a technique to extract the target speaker's voice from mixture signals using a pre-recorded enrollment utterance that characterize the voice characteristics of the target speaker.
One major difficulty of target speech extraction lies in handling variability in ``intra-speaker'' characteristics, i.e., characteristics mismatch between target speech and an enrollment utterance.
While most conventional approaches focus on improving {\it average performance} given a set of enrollment utterances, here we propose to guarantee the {\it worst performance}, which we believe is of great practical importance.
In this work, we propose an evaluation metric called worst-enrollment source-to-distortion ratio (SDR) to quantitatively measure the robustness towards enrollment variations.
We also introduce a novel training scheme that aims at directly optimizing the worst-case performance by focusing on training with difficult enrollment cases where extraction does not perform well.
In addition, we investigate the effectiveness of auxiliary speaker identification loss (SI-loss) as another way to improve robustness over enrollments.
Experimental validation reveals the effectiveness of both worst-enrollment target training and SI-loss training to improve robustness against enrollment variations, by increasing speaker discriminability.

\end{abstract}

\section{Introduction}
Although recent advances in deep learning technology improved automatic speech recognition (ASR) accuracy drastically, it remains difficult to recognize the overlapping speech of multiple speakers, which naturally occurs in realistic situations. 
Speech enhancement front-end has widely been investigated in combination with ASR back-end to recognize speech in overlapping conditions~\cite{barker2015third,barker2018fifth, comon2010handbook,li2014overview,wang2018supervised}. 

There are two major approaches explored as speech enhancement front-end.
One approach is speech separation where the mixture of speech signals are separated into signals from each source speaker\cite{luo2018tasnet,yu2017permutation,kolbaek2017multitalker,hershey2016deep}. 
Speech separation enables ASR of all speakers in the speech mixture and has shown promising performance even with a single channel setup.
However, this type of approach may suffer from the global permutation ambiguity problem i.e. arbitrary mapping between source and output signals.
The other speech enhancement approach, target speech extraction (TSE), naturally solves the permutation issue by focusing only on the signal from the target speaker~\cite{vzmolikova2019speakerbeam
,delcroix2018single,wang2018deep,delcroix2019compact,wangvoicefilter,xu2019time,ephrat2018looking,afouras2018conversation,sato2021multimodal,gu2019neural}.
TSE extracts the target speaker's speech from a mixture by using auxiliary information about the target. 
Pre-recorded target's speech is commonly used for the auxiliary information~\cite{vzmolikova2019speakerbeam
,delcroix2018single,wang2018deep,delcroix2019compact,wangvoicefilter,xu2019time}, which we call an enrollment utterance hereafter.
TSE is suitable for applications that only require recognizing a particular target speaker as those deployed in personalized devices like smartphones and smart speakers.

Despite the advantage of obtaining the target speaker-specific signal, TSE has the potential challenge of performance deviation, depending on the choices of enrollment utterances.
In general, speaker-characteristics are different between an enrollment utterance and the target speech in the mixture since the tone or intonation of speech varies depending on when we speak.
This ``intra-speaker'' speaker-characteristic variance can result in wrong extraction of the non-target speech.
This is a specific problem for TSE and not for speech separation, because TSE has to perform a more challenging task of separation and identification.

In previous works~\cite{vzmolikova2019speakerbeam}, extraction performance was typically evaluated by performing extraction with a \emph{single} enrollment utterance prepared for each mixture signal.
Thus, the performance variation with respect to enrollment choices for the same mixture was not fully considered.
However, the evaluation of the performance variation over enrollment choices is an important issue, because it is practically essential to perform TSE robustly against every possible choice of enrollment.
Especially, preventing the extraction failure caused by a specific enrollment choice is a significant challenge.

To this end, we propose a novel evaluation metric \emph{worst-enrollment source-to-distortion ratio (SDR)} as a quantitative measure of the worst performance over enrollment choices.
To evaluate the worst-enrollment SDR, we prepare multiple enrollment utterances as enrollment candidates for each mixture in evaluation data, and perform TSE with every candidate.
The worst-enrollment SDR is defined as the worst SDR value over enrollment candidates. 

In this work, we investigate two approaches to improve worst-enrollment performance.
First, to directly optimize the TSE model for the worst-enrollment SDR, we propose \emph{Worst-enrollment target training} that focuses on the difficult enrollment cases.
Specifically, worst-enrollment target training calculates the training loss for the enrollment utterance whose extraction performance is the worst within a set of several enrollment utterances selected for each mixture.
As another approach, we investigated speaker identification loss (SI-loss)~\cite{delcroix2020improving,mun2020sound} as well as its combination with the worst-enrollment target training.
The SI-loss was initially proposed to increase speaker discriminability by introducing an auxiliary speaker identification task to TSE.
They may help learn to extract more robust embeddings with less intra-speaker variance, which could also result in improving performance on worse enrollment cases, although prior works\cite{delcroix2020improving,mun2020sound} have not evaluated this.

In our results, both the proposed worst SDR-loss and SI-loss are shown to be effective to increase speaker discriminability and improve worst-enrollment SDR by 0.9 dB on average, by the combination of these approaches.
Moreover, the proposed approach drastically reduced the overall extraction failure ratio (defined as the ratio of utterances whose SDR improvement is less than 5 dB) by 34\% relative.

Our contribution is threefold:
\begin{itemize}
    \item We propose an evaluation metric of worst-enrollment SDR to quantitatively measure the worst performance over enrollment choices.
    \item We propose a novel worst-enrollment target training to directly optimize the TSE model for the worst-enrollment SDR criterion.
    \item We evaluate the effectiveness of SI-loss as well as our proposal on the worst-enrollment SDR and revealed the effectiveness of both methods for improving robustness toward enrollment variations.
\end{itemize}

\section{Conventional Method}
In this section, we describe the TSE model adopted in this work as well as the conventional way of training it.

\subsection{Overview of target speech extraction (TSE)}
In this work, we assume that a mixture signal $\Y \inRT$ is captured by a single channel microphone as $\Y = \S + \I + \N$ where $\S \inRT$, $\I \inRT$, and $\N \inRT$ denote the target speaker's speech, interfering speaker's speech, and background noise, respectively, and $T$ denotes the number of time samples.

Target speech extraction $\SE{}(\cdot)$ aims at extracting target speech from the observed mixture signal with auxiliary information about the target to inform which speaker to extract.
In this work, we use an enrollment utterance of the target speaker as the auxiliary information, which is denoted as $\Cs \in \mathbb{R}^{T_s}$ where $T_s$ is the number of samples. 
We assume that one utterance of enrollment is available in the decoding phase.
Given $\Cs$, the target signal $\Shat \inRT$ is estimated from the observed mixture as:
\vspace{-12pt}
\begin{align}
    \label{eq:se}
    \Shat = \SE(\Y, \Cs; \param)
\end{align}
where $\param$ denotes learnable parameters of the TSE network.

In this work, we adopted time-domain SpeakerBeam as a TSE architecture\cite{delcroix2020improving}. 
Time-domain SpeakerBeam is a deep neural network based speech enhancement method that is composed of two sub-networks.
\emph{Speaker embedding network} $\Embed(\cdot)$ is designed to extract the latent representation of the target speaker $\zs \in \mathbb{R}^{D}$ from the waveform signal of an enrollment utterance $\Cs$ as folows: 
\vspace{-1pt}
\begin{align}
    \label{eq:embed}
    \zs = \Embed(\Cs;\parame)
\end{align}
where $\parame$ denotes the learnable parameters of the speaker embedding network and $D$ is the dimension of the representation.
\emph{Extraction network} $\Extract(\cdot)$ extracts target speech from a mixture signal given the speaker embedding as follows:
\vspace{-2pt}
\begin{align}
    \label{eq:extract}
    \Shat = \Extract(\Y, \zs; \params)
\end{align}
where $\params$ denotes the learnable parameters of the extraction network. 

\subsection{Conventional training method}
The learnable parameters $\param = \{\parame, \params\}$ are jointly optimized to get optimal speaker representation space with respect to TSE.
The optimization target is a signal level distance between estimated target speech $\Shat$ and ground truth target speech $\S$. 
Specifically, we adopt scale-dependent SDR $\SDR(\S,\Shat)=10\log_{10}(\|\Shat\|^2/\|\S-\Shat\|^2)$ as the distance measure.
The SDR loss is denoted as $\Lsdr(\S,\Y,\Cs) = -\SDR(\S, \SE(\Y, \Cs; \param))$.
Thus, we need paired data of mixture signal, target speech, and enrollment utterance $(\S, \Y, \Cs)$ for the optimization.

To increase the variety of the combination of mixture signal and enrollment speech, multiple enrollment utterances can be prepared for each mixture in training data, and one of them is selected for training at every epoch~\cite{vzmolikova2019speakerbeam}. 
Formally, the enrollment utterance $\Cs$ is randomly or iteratively chosen from a set of candidates $\cands = \{\Csc_{1},...,\Csc_{N}\}$ where $N$ denotes the number of enrollments prepared for each mixture signal. 
In this case, the loss function $\Lcon$ is given by
\begin{align}
    \label{eq:random}
    \Lcon = \Ln\:\text{with}\: n \sim \mathcal{U}\{1,\dots,N\}
\end{align}
where $\Ln$ denotes loss calculated for $n$-th enrollment defined as $\Ln = \Lsdr(\S,\Y,\Csc_{n})$ and $\mathcal{U}\{1,\dots,N\}$ denotes the uniform distribution that takes integer values between $1$ to $N$.

\begin{table*}[t]
\centering
\caption{Data generation setup.}
\vspace{-10pt}
\label{tab:setup}
\scalebox{.95}[.95]{
\begin{tabular}{lllcccc}
\hline
 & \multicolumn{1}{c}{dataset} & \multicolumn{1}{c}{mixture type} & \begin{tabular}[c]{@{}c@{}}SIR\\ {[}dB{]}\end{tabular} & \begin{tabular}[c]{@{}c@{}}SNR\\ {[}dB{]}\end{tabular} & \begin{tabular}[c]{@{}c@{}}\#speakers\\ (train set / dev set)\end{tabular} & \begin{tabular}[c]{@{}c@{}}\#mixtures or utterances\\ (train set / dev set)\end{tabular} \\ \hline
(a) & training data for TSE & 2 speakers and noise & -5 - 5 & 0 - 20 & 777 / 49 & 50,000 / 5,000 \\
(b) & training data for ASR & 1 speaker and noise & - & 0 - 20 & 1,367 / 29 & 403,072 / 4,000 \\
(c) & evaluation data & 2 speakers and noise & -5 - 5 & 5 - 15 & 30 & 4,000 \\ \hline
\end{tabular}
}
\vspace{-8pt}
\end{table*}
\section{Strategies to improve the worst performance}
In this section, we first define the worst-enrollment SDR to quantitatively evaluate the robustness against enrollment variation, and then explain training strategies to improve it.

\subsection{Worst-enrollment SDR}
To evaluate extraction performance focusing on the robustness over enrollment variability, we propose worst-enrollment SDR metrics.
We prepare $N$ enrollment utterances $\cands$ for each mixture for evaluation, as in training.
SDR are evaluated for each extracted speech using each enrollment utterance.
$n$-th worst-enrollment SDR is defined as $n$-th worst value within $N$ enrollment choices, averaged over mixtures.
Formally, the worst-enrollment SDR $\WSDR$ is described as follows: 
\begin{align}
    \label{eq:worstsdr}
    \WSDR = \min_{n \in \cands} \SDR(\S, \SE(\Y, \Csc_{n}; \param)).
\end{align}
The same definition holds for other metrics than SDR such as character error rate (CER).

\subsection{Worst-enrollment target training}
In the previous training method, enrollment utterances are sampled from candidates randomly or iteratively as in Eq.~\eqref{eq:random}. 
Thus, all prepared enrollments are treated equally in training.
However, it may not be optimal in terms of the worst-enrollment SDR.
To directly optimize TSE performance on worst-enrollment SDR, we propose worst-enrollment target training that trains the model  with weight placed on difficult enrollments.

For worst-enrollment target training, $K(\leq N)$ enrollment utterances are randomly sampled from enrollment candidates $\cands$ to make the subset $\subcands$ for each mixture at each epoch.
In \emph{hard} version of worst-enrollment target training, the loss function $\Lwh$ is calculated as the maximum, thus worst, value of minus SDR loss over $K$ enrollments as follows:
\begin{align}
    \label{eq:worsthard}
    \Lwh = \max_{n\in \subcands} \Ln.
\end{align}
The sampling of $\subcands$ from $\cands$ is introduced to reduce computational cost in model forwarding. It also prevents focusing only on a certain enrollment utterance and decreasing enrollment variation in the training.

We also propose a \emph{soft} version of the worst-enrollment target training to prevent the model from excessively adapting to difficult cases e.g. outlier enrollments that is too noisy or different from the typical speaking style of the target speaker.
The loss function $\Lws$ softly weight difficult cases over $K$ enrollments using a softmax function with temperature as follows:
\vspace{-8pt}
\begin{align}
    \label{eq:worstsoft}
    \Lws = \sum_{n \in \subcands} w_{n} \, \Ln \\
    w_{n} = \frac{\exp(\Ln/\tau)}{\sum_{n' \in \subcands} \exp(\Lnd/\tau)}
\end{align}
where $w_{n}$ denotes the weights and $\tau$ denotes the temperature parameter of the softmax function, respectively. 
The temperature parameter controls the degree of smoothing over candidates.
The hard version can be seen as $\tau \to 0$ limit.

\subsection{Auxiliary speaker identification loss}
Auxiliary SI-loss~\cite{delcroix2020improving} is an approach that was proposed to help speaker embedding network learn to extract more invariant latent expression to intra-speaker variance.
It may also improve robustness toward enrollment variation by better handling difficult enrollment utterances e.g. those close to speaker boundaries.
In SI-loss training, an auxiliary speaker identification task given a speaker embedding $\zs$ is introduced to improve speaker discriminability. 
The multitask loss for SI-loss training $\Lsi$ is given by,
\begin{align}
    \label{eq:multitask}
    \Lsi = \Lsdr + \alpha\,\CE(\one, \softmax(\W\zs))
\end{align}
where $\one$ denotes a one-hot vector representing the target speaker ID, $\W$ denotes an additional learnable projection matrix, and $\CE(\cdot)$ and $\softmax(\cdot)$ are cross entropy and softmax function, respectively. $\alpha$ is a multitask weight parameter.

We also investigated the combination of the hard version of worst-enrollment target training and auxiliary SI-loss.
For its training, we replaced the $\Lsdr$ in Eq.~\eqref{eq:multitask} with $\Lwh$ and calculate CE loss only for the worst enrollment within $\subcands$ in terms of the SDR loss.

\begin{figure}[bt]
 \begin{center}
  \includegraphics[width=1\hsize]{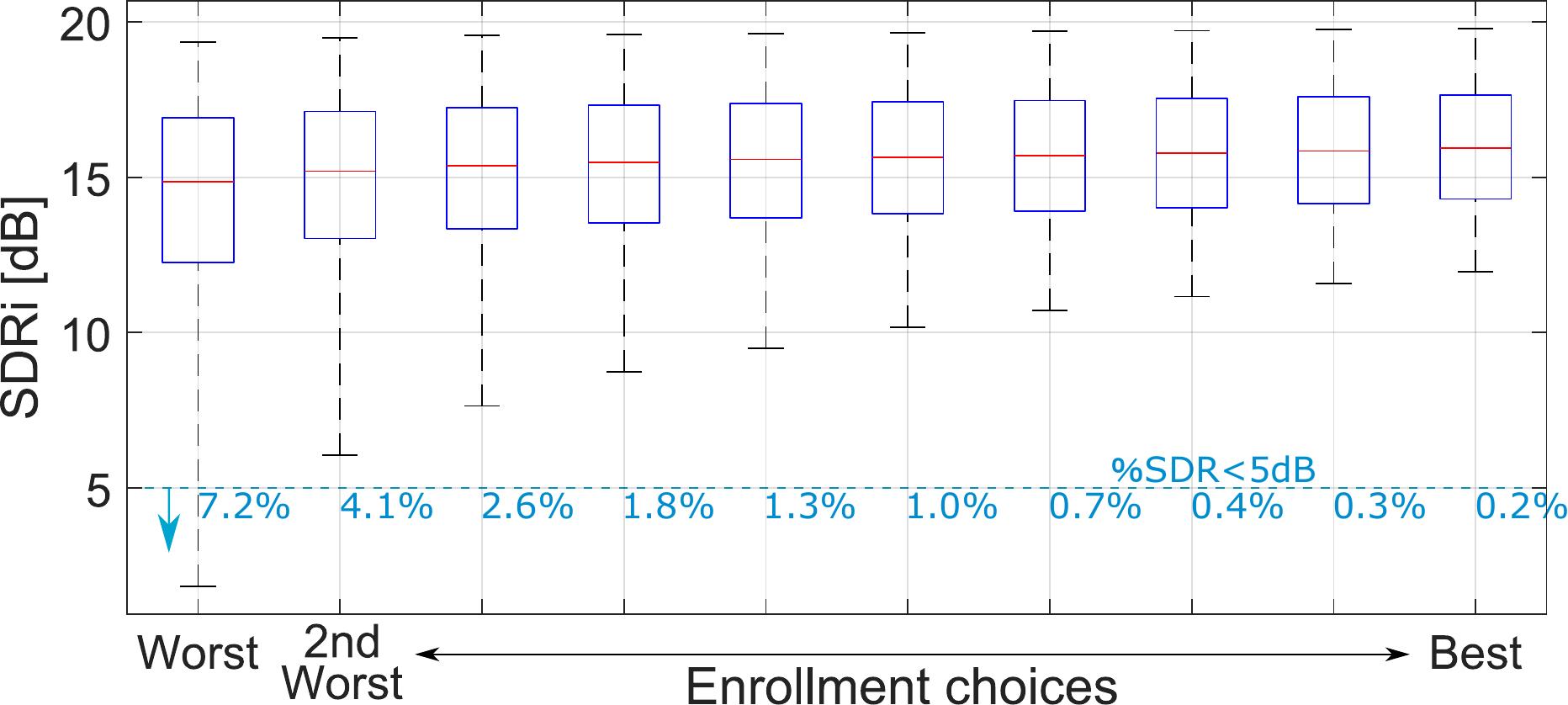}
 \end{center}
 \vspace{-12pt}
 \caption{Extraction performance as a function of the enrollment, from worst to best, for the baseline TSE system. The number in blue indicate the failure rate, i.e. the ratio of samples whose SDRi is below 5dB. 
 Each box plot indicates 5th, 25th, 50th, 75th, 95th percentile values aggregated over 4,000 mixtures in evaluation dataset. }
 \vspace{-12pt}
 \label{fig:distribution}
\end{figure}

\begin{table*}[t]
\centering
\caption{Performance of each system for worst, 2nd-worst and best enrollment choices, as well as average over enrollments. Both worst-enrollment target training and auxiliary SI-loss improve the worst-enrollment performance in terms of SDRi and CER.}
\vspace{-8pt}
\label{tab:main}
\scalebox{0.96}[0.96]{
\begin{tabular}{l|cccc|cc|ccc}
\hline
 \multicolumn{1}{c|}{Metrics} & \multicolumn{4}{c|}{SDRi  {[}dB{]} (↑)} & \multicolumn{2}{c|}{CER {[}\%{]} (↓)} & \multicolumn{3}{c}{\begin{tabular}[c]{@{}c@{}}Failure ratio {[}\%{]} (↓)\\ (SDRi \textless 5dB)\end{tabular}} \\
 \multicolumn{1}{c|}{Enrollment choices} & mean & worst & \begin{tabular}[c]{@{}c@{}}2nd\\ worst\end{tabular} & best & mean & worst & mean & worst & best \\ \hline
Conventional & 15.1±3.8 & 13.3 & 14.3 & \textbf{15.9} & 13.3 & 19.6 & 2.0 & 7.7 & 0.2 \\ \hline
Worst loss (hard) & 15.2±3.6 & 14.0 & \textbf{14.8} & 15.8 & 13.1 & 18.1 & 1.4 & 4.8 & 0.2 \\
Worst loss (soft) & 15.2±3.4 & 14.1 & \textbf{14.8} & 15.8 & 13.1 & 18.2 & \textbf{1.3} & 4.5 & 0.2 \\
SI-loss & \textbf{15.3±3.5} & 14.1 & \textbf{14.8} & \textbf{15.9} & \textbf{13.0} & 17.6 & \textbf{1.3} & 4.7 & \textbf{0.1} \\
Worst loss (hard) + SI-loss & \textbf{15.3±3.5} & \textbf{14.2} & \textbf{14.8} & 15.8 & \textbf{13.0} & \textbf{17.5} & \textbf{1.3} & \textbf{4.4} & 0.2 \\ \hline
\end{tabular}
}
\vspace{-10pt}
\end{table*}

\begin{figure}[t]
 \begin{center}
  \includegraphics[width=0.98\hsize]{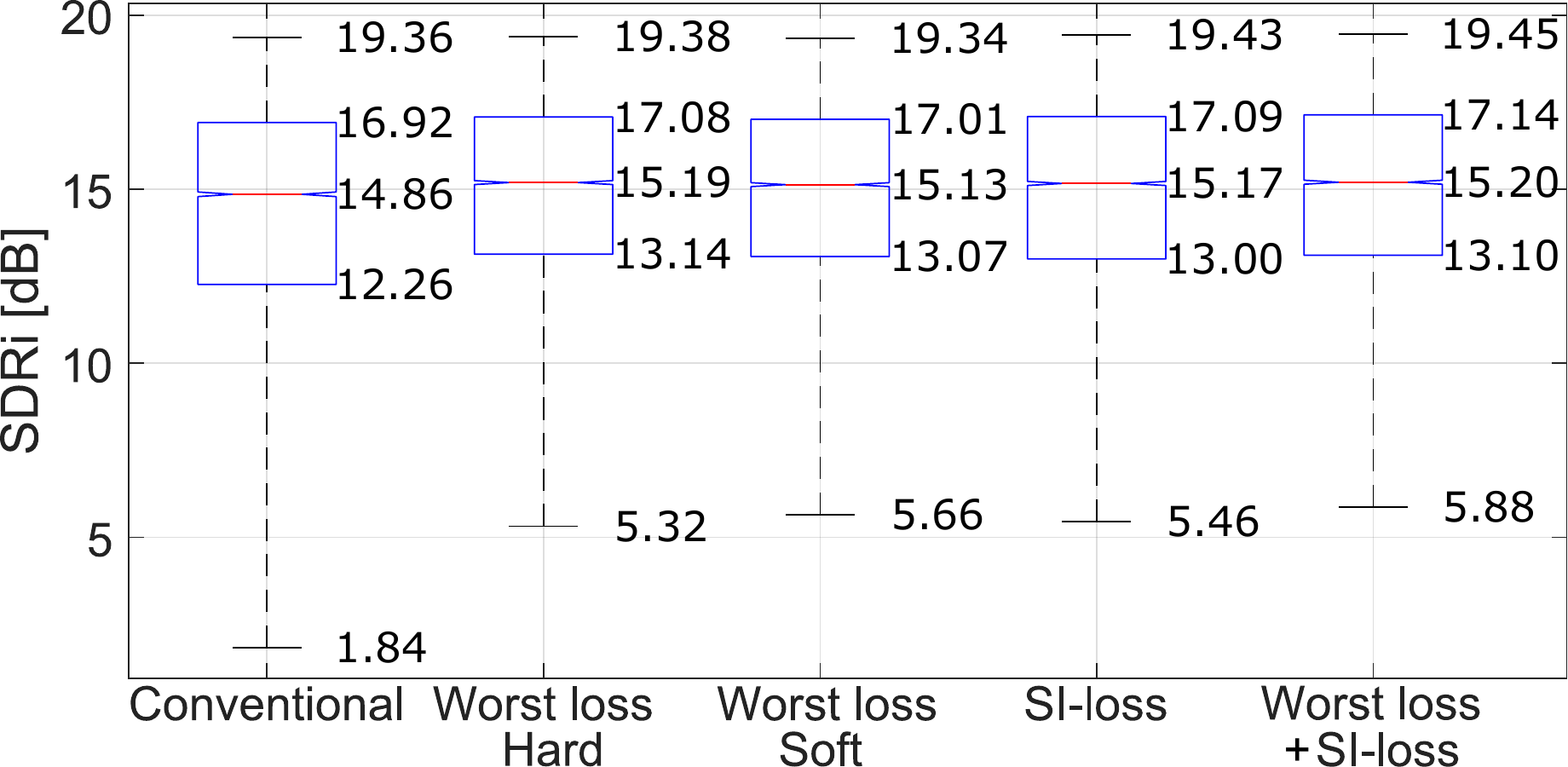}
 \end{center}
 \vspace{-12pt}
 \caption{Worst-enrollment SDRi for conventional and proposed methods. Each box plot indicates 5th, 25th, 50th, 75th, 95th percentile values. All the training strategies improve the lower limit performance.}
\vspace{-12pt}
 \label{fig:boxplot}
\end{figure}
\section{Experiments}
\subsection{Experimental Setup}
\subsubsection{Evaluation details}
Five systems were prepared for evaluation: a conventional system trained with loss function in Eq.~\eqref{eq:random}, a system trained with hard worst-enrollment loss in Eq.~\eqref{eq:worsthard}, a system trained with soft worst-enrollment loss in Eq.~\eqref{eq:worstsoft}, a system trained with an auxiliary SI-loss in Eq.~\eqref{eq:multitask}, and a system trained with the combination of the worst-enrollment training and an auxiliary SI-loss.
We evaluated the performance in terms of SDR improvement (SDRi), character error rate (CER), and the failure ratio.
Here we define the failure ratio as the ratio of data whose SDRi was less than 5 dB in evaluation data. 
We computed scale-dependent SDR for evaluation using BSS Eval toolbox~\cite{vincent2006performance}.

\subsubsection{Dataset}
All the training and evaluation were performed on simulated mixtures generated from speech recordings from the Corpus of Spontaneous Japanese (CSJ)~\cite{maekawa2003corpus} and noise recordings from the CHiME-4 corpus\cite{chime4}, sampled at 16 kHz.
Table \ref{tab:setup} shows the data generation setup for training and evaluation. 
Signal-to-interference ratio (SIR) and signal-to-noise ratio (SNR) in the table stands for the energy ratio between the signals from target and interfering speakers, and between the speech mixture without noise and the noise. 
For training and evaluation of the TSE model, we prepared $N=10$ auxiliary utterance candidates for each mixture. 
Candidates were selected randomly from the target speaker's utterances except those used as target speech and shorter than two seconds.

\subsubsection{System Configuration and Training Procedure}
We adopted time-domain SpeakerBeam architecture for TSE \cite{delcroix2020improving}. 
We used blocks of eight stacked 1-D convolution layers as proposed in \cite{luo2019conv} for the auxiliary and extraction networks, repeated three times for the extraction network. The other parameters of the model follow our prior work \cite{delcroix2020improving} and are omitted here because of space limitations.
For ASR evaluation, we adopted transformer-based encoder-decoder ASR model with connectionist temporal classification objective~\cite{karita2020ctctransformer} trained according to the recipe of the  ESPnet open-source toolkit~\cite{watanabe2018espnet}. 

For optimization, we adopted the Adam optimizer\cite{kingma2014adam}. 
The initial learning rate was set as 5e-4 and was halved if the loss on the development set did not decrease for 3 epochs. 
The models were trained for 120 epochs. SI-loss was adopted from scratch, whereas worst-enrollment target training was adopted from 100 epochs.
The best-performing model based on a development set was chosen for evaluation.
For worst-enrollment target training, the number of selected enrollments from candidates was $K=3$ for both hard and soft versions.
The temperature parameter in soft worst-enrollment target training was set as $\tau = 2.0$.
Multitask weight in auxiliary SI-loss training was $\alpha = 1.0$.

\subsection{Results}
\subsubsection{Effect of enrollment variations on the performance}
In this section, we discuss how much enrollment choices affect performance.
Fig.~\ref{fig:distribution} shows the performance distribution of $n$-th worst-enrollment SDRi obtained with the conventional system. 
The ratio of data whose SDRi was less than 5 dB is also presented as the failure ratio.
The figure shows that extraction performance varies depending on enrollment choices even though each plot shares the same mixtures.
Especially, the 5 and 25 percentile values differ greatly between $n$-th worst-enrollment SDRi. 
In other words, enrollment choices affect the lower bound performance (failure cases) rather than the upper bound performance.
In addition, the failure ratio of 7.2 \% for the worst-enrollment SDRi is much greater than other enrollment choices. 
This indicates that enrollment choices have a high possibility of determining whether TSE succeeds.
Thus, the robustness toward enrollment choices is an issue to be addressed.

\subsubsection{Evaluation on speech extraction performance}
Here we discuss the performance of each training method.
Table~\ref{tab:main} shows SDRi, CER and failure ratio of each system with worst, 2nd-worst and best-enrollment choices as well as averaged values over enrollments.
The table shows that both worst-enrollment loss and SI-loss improved performance for worst and 2nd-worst enrollment choices.
Hard/soft worst-enrollment loss training, SI-loss training, and the combination of these two approaches improved worst-enrollment SDR by 0.7, 0.8, 0.8, and 0.9 dB and CER by 1.5, 1.4, 2.0, and 2.1 \% points, respectively.
As a deeper evaluation of worst-enrollment performance, Fig.~\ref{fig:boxplot} shows the performance distribution of the worst-enrollment SDRi for each system. 
Hard/soft worst-enrollment loss training, SI-loss training, and the combination of them improved 5 percentile values largely from 1.84 dB to 5.32, 5.66, 5.46, and 5.88 dB, respectively.
From the observations above, these strategies successfully improve robustness towards enrollment variation by especially improving lower bound performances in worst enrollment cases.

As for the failure ratio of each system, as shown in table \ref{tab:main}, all the training methods reduce the failure ratio, especially for the worst enrollment cases.
The failure ratio was also improved on average over enrollments, by 34 \% relative from conventional training method with the combination of hard worst-enrollment loss training and auxiliary SI-loss.
This seems to be because the failure of extraction most often occurs in worst enrollments cases as seen in Fig.~\ref{fig:distribution}.
From this observation, we can say that these approaches contribute to reducing the overall failure that often occurs for difficult enrollment utterances by improving the worst performance.

\subsubsection{Analysis on speaker discriminability}
To further understand the effect of these two types of strategies, we conducted analysis on the speaker discriminability of each system.
Table~\ref{tab:speaker} shows the speaker discriminability evaluated with between-class within-class variance ratio of speaker embeddings $\zs$~\cite{bishop2006pattern,dehak2010front}.
The training with SI-loss improves the speaker variance ratio as expected, by introducing an auxiliary speaker identification task.
It should be noted that worst-enrolment loss also improves the speaker discriminability, and the combination of the worst-enrollment loss with auxiliary SI-loss shows the largest improvement.
This may explain the superior performance of the combination of these approaches to the worst-enrollment performance shown in Table~\ref{tab:main}.
The evaluation with a more difficult setup, where the speaker characteristics of the interfering speaker are more similar to those of the target speaker e.g. recordings from the speakers belonging to the same family will be a part of our future work.

\begin{table}[t]
\vspace{-5pt}
\centering
\caption{Intra-speaker inter-speaker variance ratio of speaker embeddings. Higher values indicate higher speaker discriminability learned by the embedding network.}
\vspace{-6pt}
\label{tab:speaker}
\scalebox{.96}[.96]{
\begin{tabular}{l|c}
\hline
\multicolumn{1}{c|}{} & variance ratio \\ \hline
Conventional & 2.96 \\ \hline
Worst loss (soft) & 3.35 \\
SI-loss & 3.49 \\
Worst loss (hard) + SI-loss & \textbf{3.61} \\ \hline
\end{tabular}
}
\vspace{-14pt}
\end{table}



\section{Conclusion}
In this work, we focused on the performance variance of TSE depending on the choice of enrollment utterances.
We introduced a new performance metric, worst-enrollment SDRi, to quantitatively evaluate the robustness toward enrollment variability. 
To directly optimize the network for worst-enrollment SDRi, we proposed worst-enrollment target training that trains the model with placing weight on difficult enrollments.
We also evaluated auxiliary speaker identification loss in terms of worst-enrollment performance, which was previously proposed for improving speaker discriminability.
Experimental validation showed the effectiveness of both approaches to increase the robustness toward enrollment variability.

\pagebreak

\bibliographystyle{IEEEtran}

\bibliography{mybib}
\end{document}